\documentclass[12pt,letterpaper]{JHEP3}
\usepackage{graphicx}

\usepackage{color}

\usepackage{soul}

\usepackage{cite}

\graphicspath{{Figures/}}

\title{Coherent Cascade Conjecture for
Collapsing Solutions in Global AdS}

\author{Ben Freivogel\footnote{benfreivogel@gmail.com} \ and I-Sheng Yang\footnote{isheng.yang@gmail.com}\\
$^*$ GRAPPA and ITFA, Institute of Physics, Universiteit van Amsterdam, \\
Science Park 904, 1090 GL Amsterdam, Netherlands \\ \ \\
$^\dagger$ Perimeter Institute of Theoretical Physics, \\
31 Caroline Street North, Waterloo, ON N2L 2Y5, Canada, \\ and, \\
Canadian Center of Theoretical Astrophysics, \\
60 St George St, Toronto, ON M5S 3H8, Canada
}

\abstract{
We analyze the gravitational dynamics of a classical scalar field coupled to gravity in asymptotically AdS spacetime, which leads to black hole formation on the shortest nonlinear time scale for some initial conditions. We show that the observed collapse cannot be described by the well-known process of a random-phase cascade in the theory of weak turbulence. This implies that the dynamics on this time scale is highly sensitive to the phases of modes. We explore the alternative possibility of a coherent phase cascade and analytically find stationary solutions with completely coherent phases and power-law energy spectra. We show that these power-law spectra lead to diverging geometric backreaction, which is the likely precursor to black hole formation. In 4+1 dimensions, our stationary solution has the same power law energy spectrum as the final state right before collapse observed in numerical simulations. We conjecture that our stationary solutions describe the system shortly before collapse in other dimensions, and predict the energy spectrum.
}

\begin{document}

\section{Introduction and Summary}

The nonlinear stability of global Anti-deSitter (AdS) spacetime has received increasing attention in the past few years. The other two maximally symmetric spacetimes, Minkowski and deSitter, have been shown to be stable\cite{ChrKla93,Fri86}. Intuitively speaking, the main difference is that in those cases, energy can escape to infinity. Global AdS spacetime, however, comes naturally with reflecting boundary conditions. Any initial perturbation, no matter how small, is confined to gravitationally interact with itself, effectively in a finite region, forever. Therefore it is more likely for the energy distribution to become highly uneven and back-react strongly on the metric. The long-term outcome of such nonlinear dynamics is difficult to predict, which resulted in the richness of the current AdS (in)stability problem\cite{BizRos11,DiaHor11,Lie12,deOPan12,DiaHor12,BucLie13,MalRos13,MalRos13a,AllFir14,
MalRos14,BalBuc14,OkaCar14,HorSan14,Mal12,BucLeh12,BizJal13,Jal13,BizRos14,
DFLY14,CraEvn14,CraEvn14a,BucGre14,BizMal15,BucGre15,BucGre14a,BasKri15,Yan15,DimYan15,GreMai15,EvnKri15,CraEvn15a,CraEvn15b}.

Four years ago, Bizon and Rostworowski presented interesting numerical results that spurred recent developments \cite{BizRos11}. They showed that initial perturbations with a small amplitude $\epsilon$ collapse to a black hole on the nonlinear time scale,  $T \sim \epsilon^{-2}$. This result is very interesting. Perturbation theory guarantees that the instability cannot develop at any time scale shorter than $T\sim\epsilon^{-2}$, so the BR result suggests that global AdS may be unstable at the shortest possible time scale allowed by the dynamics.  Furthermore, an instability can lead to many different deviations from empty AdS, and it has no particular reason to directly become a black hole, but the BR result suggests that black hole formation may be the generic outcome.

Black hole formation is a natural endpoint of the dynamics, since in one way of taking the classical limit, the black hole is the equilibrium configuration in the microcanonical ensemble \cite{BanDou98,DFLY14}.  However, there is certainly no guarantee that the system will equilibrate on this short time scale. In addition, a generic expectation is that as the interaction strength decreases, the ergodic region of phase space decreases in size. This result is proven by the KAM theorem for a wide class of systems \cite{KAM}, but the assumptions of the theorem are not met here. So it is interesting to ask how generic black hole formation is, particularly on the nonlinear time scale. This question is dual to the question of how efficient thermalization is in a large N conformal field theory on a spatial sphere of radius $R$, at energies in the range
\begin{equation}
R^{-1} \ll E \ll {N^2 \over R}~.
\end{equation}

Further work showed that although some initial conditions do lead to collapse, black hole formation is not the only possible behavior at the $T\sim\epsilon^{-2}$ time scale. In particular, there exist open sets of initial conditions that avoid collapse on this time scale. Such a strong conclusion is based on two cornerstones: First, the discovery of ``islands of stability"  \cite{DiaHor12,BucLie13,MalRos13a}, and study of the phase-coherent dynamics of these non-collapsing solutions in the two-time formalism \cite{GreMai15,CraEvn15b}; and second, the conditional reliability of the two-time formalism and rescaling symmetry that guarantees that these solutions survive in the small amplitude limit, at the $\epsilon^{-2}$ time scale \cite{DimYan15}.

In this paper, we take a step towards a similar understanding for collapsing solutions. Some evidences showed that open sets of stable solutions are anchored special solutions with stationary exponential energy spectra \cite{GreMai15,CraEvn15b}. We will present possibly analogous special solutions for collapsing solutions: stationary solutions with power-law spectra which causes the right amount of geometric backreaction for black hole formation.

First of all, we have to point out a  confusion caused by how the term ``weak turbulence'' is used in the AdS (in)stability problem.  As described in many pioneering works \cite{BizRos11,DiaHor11,MalRos13,Biz13,BucGre14}, weak gravitational interactions between the AdS eigenstates of linearized perturbations, expanded to the first nonlinear order, lead to quartic couplings between the modes. This type of system has been extensively studied  in the theory of weak turbulence. However, it is misguided to claim that the Kolmogorov-Zakharov power law in weak turbulence \cite{Kolmogorov,WeakTurbulence} explains the power law spectrum seen in the collapsing AdS solutions \cite{deOPan12}. 

In the weak turbulence context, the equations of motion are solved under the random phase ansatz. The phases of eigenstates are assumed to be randomly distributed, thus any phase-dependent effect averages to zero. The phase information is therefore totally discarded, and the amplitudes of eigenstates are described by a phase-independent dynamics.  A basic result under the random phase ansatz is that for a system with quartic interactions, there is {\it no energy transfer} between the modes on the $\epsilon^{-2}$ time scale; the nontrivial dynamics occurs on the longer $\epsilon^{-4}$ time scale \cite{Kolmogorov, WeakTurbulence} \footnote{Since \cite{Kolmogorov, WeakTurbulence} are quite technical and contains a lot of other information, it may not be straightforward to understand this point directly by reading it. We sketch a simple derivation in Appendix \ref{sec-random} to show the readers how random-phase ansatz kills any dynamics in the $\epsilon^{-2}$ time scale.}.
If one really takes the full analogy to weak turbulence to solve the dynamics of AdS eigenstates, the Kolmogorov-Zakharov power law will be attained at the $\epsilon^{-4}$ time scale, which cannot explain the observed black hole formation in the shorter, $\epsilon^{-2}$ time scale.
Therefore, in order to understand the results of simulations, we must pursue an analytical strategy that keeps track of the phases and their relevance in the dynamics.

Keeping track of the phases makes the problem considerably harder. Fortunately, experience tells us that even the simplest possible solution can provide a lot of insight. Recall that the sets of stable solutions, sometimes called stability islands, are anchored on special ``quasi-periodic solutions'', namely exactly stationary solutions, with coherent phases and exponential spectrum, in the two-time approximation \cite{BalBuc14, BucGre14, GreMai15, CraEvn15b}. We will show that the two-time approximation contains another type of exactly stationary and coherent solution which likely play the same role for collapsing solutions. Although stationary solutions do not really evolve, their possible forms are highly constrained, providing important information of the dynamics. Finding them is often the first step toward understanding other solutions with similar properties \cite{GreMai15,CraEvn15b}. The solution we find has the following properties:
\begin{itemize}
\item Instead of an exponential spectrum, these solutions have a power law spectrum.\footnote{The power law solutions evaded earlier solution searches \cite{GreMai15,CraEvn15b} either due to some starting assumption that excluded power laws from the very beginning, or due to removing solutions by hand if they run into a UV cut-off, which power law solutions usually do.}
\item Within the two-time approximation, these solutions are protected by the rescaling symmetry, thus also persist in the $\epsilon\rightarrow0$ limit.
\item These solutions come with specific power laws, $E\sim w^{2-d}$ as a function of the frequency $\omega$, which agree with the extensive numerical observations in $d=4$.
\item The back-reaction from these coherent power law solutions is strong enough to give finite deviation from empty AdS even in the $\epsilon\rightarrow0$ limit. In particular, the deviation is suggestive of black hole formation.
\end{itemize} 

The backreaction calculation is not difficult, but is has been neglected in the recent literature. Although the possibility of black hole formation largely motivated the current developments, much recent work has focused on the scalar field spectrum without establishing an explicit link to the actual geometric backreaction. Such a link is necessary to establish that AdS space is indeed unstable, and to understand the outcome of instability.

In Sec. \ref{sec-kinematics}, we calculate the relation between the scalar field power spectrum and the geometric backreaction it causes, with particular emphasis on diagnosing whether the backreaction is singular, and when it is, whether it suggests a black hole, or some other type of singularity. We find that the phase coherence between modes strongly affects the backreaction. We compare the fully coherent and the fully incoherent cases and find that phase coherence leads to stronger backreaction from the same power law. In particular, we observe that the power laws found in numerical simulations are probably insufficient to imply black hole formation if the phases are incoherent. On the other hand, the same power laws with coherent phases strongly suggest black hole formation.

In Sec.\ref{sec-lock}, we derive the stationary coherent power law solutions from the recently reported scaling behavior of the coupling coefficients in \cite{CraEvn15a}. Phase-coherent dynamics explicitly predicts these power law in $AdS_{d+1}$ with arbitrary $d>3$. The energy per mode as a function of frequency is given by
\begin{equation}
E \sim \omega^{2-d}~,
\end{equation}
where $d$ is the spatial dimension of the bulk theory. The $d=4$ result exactly agrees with many numerical observations, while higher $d$ can be checked in the future. The analysis in $d=3$ is more subtle due to anomalies in the scaling and a possible dependence on the UV cut-off. Our result there is in some tension with the scaling reported in simulations, and we specifically point out possible causes to study in the future.

We do not have a precise mathematical argument relating our stationary solutions to the dynamical collapse.
However, we expect that our time-independent power law is a good description of the dynamics for a range of wavelengths that are well-separated from the long-wavelength scale, where the energy is initially injected, and the (time-dependent) UV scale where the modes have not yet been populated.

%Finally, we remind the reader that despite the agreement with numerical collapsing solutions, 
Finally, we explore possible relations between the stationary power law solution we found and the actual dynamical evolution from initial data into high modes. It is likely too na\"ive to imagine all collapsing solutions as ``approaching'' these stationary coherent power law solutions. Recall that typical stable solutions do not approach quasi-periodic solutions either. They only evolve around orbits which seem to center on the quasi-periodic solutions \cite{GreMai15}. The corresponding behavior of a typical unstable solution is likely more complicated. We suggest a possible first step in this direction by noticing that two-mode initial data seems to be particularly prone to collapse in existing numerical results. In Sec.\ref{sec-seed} we analytically derive that indeed two-mode initial data directly leads to an initially phase-coherent energy cascade. Although this is only valid for time less than $\epsilon^{-2}$, it might be an interesting starting point. For example,  three-mode initial data does not have the same simple phase coherent structure at early time. One can then numerically study the fate of three-mode initial data to see whether there are significant differences.

\section{Gravitational backreaction of coherent and incoherent power laws}
\label{sec-kinematics}

In this section, we analyze the gravitational backreaction when a number of modes are turned on. We find qualitatively different behaviors when the phases of the modes are taken to be coherent than when they are incoherent\footnote{A more precise definition of coherence will become clear later.}. We particularly focus on a power law spectrum of amplitudes, since this is the case that is seen in numerical simulations. Our convention is to parametrize the amplitudes as a power law of order $-\alpha$, namely $A_n \sim n^{-\alpha}$. The allowed frequencies for a massless field in global AdS are discrete, with $\omega_n = 2n + d$ in units of the AdS radius, so a power law in frequency is equivalent to a power law in mode number $n$. In the following, we work with the mode number $n$.  Some other papers in this field use the corresponding energy spectrum, which will be $E_n\sim n^{2-2\alpha}$. First we present out results in this chart.
\begin{center}
\begin{tabular}{| c | c | c | c | c |} 
\hline
       & regular & naked curvature & naked redshift & black hole \\ 
d=3 &               & singularity             & singularity         &  \\ 
\hline
incoherent phases &  
$\alpha>\frac{5}{2}$ & 
$\frac{5}{2}\geq\alpha>\frac{3}{2}$ & 
$\alpha \leq \frac{3}{2}$ & never \\
\hline 
coherent phases &
$\alpha>3$  &
$3\geq\alpha>2$ &
$\alpha = $2 & 
$\alpha<2$ \\
\hline 
$d>3$ &               &              &          &  \\ 
\hline
incoherent phases &  
$\alpha>\frac{d+2}{2}$ & 
$\frac{d+2}{2}\geq\alpha>\frac{d}{2}$ & 
$\alpha = \frac{d}{2}$ & 
$\alpha < \frac{d}{2}$ \\
\hline 
coherent phases &
$\alpha>\frac{d+3}{2}$  &
$\frac{d+3}{2}\geq\alpha>\frac{d+1}{2}$ &
$\alpha = \frac{d+1}{2}$ & 
$\alpha<\frac{d+1}{2}$ \\
\hline 
\end{tabular}
\end{center}
We can see that in $(3+1)$ dimensions, independent of what power law we have, incoherent phases can never correspond to black hole. Thus an observation of black hole formation together with any power law implies phase coherence. In $(4+1)$ dimensions, the numerical collapses reported values of $\alpha$ very close to 2. Since such a value is right at the edge for incoherent phases, one cannot be as conclusive. However, coherent phases still leave less doubt about the connection between this power law and black hole formation.

Since we are calculating the backreaction using the leading order expansion, and the last three columns in this chart actually correspond to diverging backreaction that invalidates the expansion, we should explain their physical meanings more carefully. 

These power law solutions are always well defined as a dynamically evolving set of harmonic oscillators described by the two-time formalism \cite{BalBuc14,CraEvn14a,BizMal15} \footnote{In a forthcoming publication, we will explain more explicitly how the oscillating singularity shown in \cite{BizMal15} disappears in the boundary time gauge.}, which approximates the actual gravity evolution before back-reaction reaches order one. So what we actually calculate is a ``fictitious back-reaction'' which approximates the actual back-reaction before it reaches order one.  If we start with some initial conditions with small back-reactions, evolve them with the two-time formalism and they reach any of the diverging power laws, then before that time, the actual back-reaction does become order one. Reaching order one back-reaction is already sufficient to show an instability.  Evolving toward a diverging fictitious back-reaction in the two-time formalism then guarantees an order one back-reaction for the gravity evolution with an arbitrarily small initial amplitude. In particular, the form of the diverging fictitious back-reaction represent the form of the actual back-reaction when it reaches order one. So one can ask whether such form is similar to a Schwarzschild metric or not, which can be a good sign of what type of large back-reaction it will approach afterward.

\subsection{Geometric Deviation}
\label{sec-norm}

% When the backreaction from the scalar field is small, the norm can be naturally defined as the perturbation it causes in the metric. First of all, we need to fix a gauge, which is simple in the case with spherical symmetry. We can demand that the radial coordinate is always the area radius of the $(d-1)$-dimensional sphere, and then its orthogonal direction uniquely defines the time coordinate, up to an overall rescaling. We then fix that at the asymptotic boundary. The norm of a scalar field configuration $\phi$ is then defined to be the norm of the change in the $t-r$ metric.
% \begin{equation}
% \|\phi(t)\| \equiv \rm{Max}
% \left(\sqrt{\delta g_{tt}^2(t') + \delta g_{rr}^2(t')},~0\leq r \leq \infty,~t\leq t' \leq t+2\pi\right)~.
% \label{eq-norm}
% \end{equation}

With spherical symmetry, we can demand to always put the metric into a standard form that easy to compare with empty AdS \footnote{The AdS radius is set to 1 for convenience throughout this paper.},
\begin{equation}
ds^2 = -(1+r^2)dt^2 + \frac{dr^2}{1+r^2} + r^2 d\Omega_{d-1}^2~.
\end{equation}
We can fix the gauge for the perturbations by enforcing that $r$ is always the area radius of the $(d-1)$ sphere, $g_{tt}$ approaches the empty AdS value at $r\rightarrow\infty$, and the off-diagonal term $g_{tr} = 0$, leaving only the physical quantities  $\delta g_{tt}$ and $\delta g_{rr}$. In particular, in defining which solutions have large backreaction, we care about their maximum values in all space 
%(but that will trivially always be at $r=0$)
 and in all time within one AdS period.\footnote{The reason for scanning through one AdS time was explained in \cite{DimYan15}. At time $t$, a large and dilute shell may be tuned to converge at $r=0$ and form a black hole within one AdS time, which is not the long time scale instability we are studying. Thus such a finely tuned dilute shell, although only modifies the metric mildly at that moment, must already mean a large geometric backreaction for our purpose.}

In a more general treatment without spherical symmetry, defining which solutions have large backreaction is nontrivial due to the freedom of gauge choice. It could be defined through a max-min scheme: scanning through the entire spacetime for the largest $\delta g_{\mu\nu}$, then all all possible gauge choices to minimize it. We will leave such an endeavor to future work.

We are interested in the case that the total energy is small and approaching zero. Thus the only possibility to have a large backreaction is to focus the energy into a small region, much smaller than the AdS radius, which was set to one. Such a region can be effectively described locally by Minkowski space, for which the perturbative expansion of small backreaction is well-known,
\begin{eqnarray}
ds^2 = -\left(1+\frac{2M}{r^{d-2}}+4V\right)dt^2 + \left(1+\frac{2M}{r^{d-2}}\right)dr^2 + r^2 d\Omega_{d-1}^2~.
\label{eq-Mink}
\end{eqnarray}
Here $M$ and $V$ are the usual definition of the enclosed mass and gravitational potential,
\begin{eqnarray}
M(r) &=& \int_0^{r}~\frac{\dot\phi^2+\phi'^2}{2}~dr'~,\\
V(r) &=& -\int_{r}^\infty~(d-2)\frac{M(r')}{r'^{d-1}}~dr'~.
\end{eqnarray}

Using this approximation, we have
\begin{equation}
\delta g_{tt} = \frac{2M(r)}{r^{d-2}}+4V(r)~, \ \ \ \ \ \delta g_{rr} = \frac{2M(r)}{r^{d-2}}~.
\end{equation}
This not only allows us to estimate whether backreaction is large, but we can further describe the physics of the backreaction, for example whether it is approaching a black hole, which requires $\delta g_{tt} \approx -\delta g_{rr}$. Some may worried that we are only keeping the leading order in the metric deviation, while higher order terms always become important in an actual black hole formation. We remind the reader that the formal mathematical definition of an instability is whether {\it an infinitesimal initial perturbation leads to a finite perturbation}, in which {\it a finite perturbation} can be still small and well-approximated by the leading order term. It is true that strictly speaking, developing some small but finite $\delta g_{tt} \approx -\delta g_{rr}$ does not guarantee black hole formation. However, it is clearly different and more suggestive of such a possibility, compared to situations in which $\delta g_{tt}$ is very different from $-\delta g_{rr}$. 

It would be very interesting to extend our treatment beyond linearized backreaction. However, this would require a number of nontrivial steps, such as defining the modes of the scalar field in the presence of nonlinear metric perturbations.

\subsection{Single Mode}

Following the previous section, we can calculate the geometric backreaction of any field configuration. As a warm-up exercise, we first consider the situation where all energy is in one eigenstate, $\phi(t,x) = A_n e_n(x) \cos w_nt$ and $E = w_n^2 A_n^2$. The energy density $\rho$ is given by
\begin{eqnarray}
\rho_n(r) \approx w_n^2 A_n^2 e_n^2 
&\sim& E n^{d-1} \ \ \ \ \ {\rm for} \ \ r\leq n^{-1}~, 
\label{eq-rhoin} \\
&\sim& \frac{E}{r^{d-1}} \ \ \ \ \ {\rm for} \ \ 
n^{-1}<r<1~.
\label{eq-rhoout}
\end{eqnarray}
This directly follows from the large $n$, small $r$ behavior of the eigenfunctions $e_n(r)$ in Eq.~(\ref{eq-eigen}). Actually, instead of going through the hypergeometric function for the actual $e_n$, one can easily derive this energy distribution with the following physical intuition. Spherically symmetric eigenstates are basically standing waves from the superpositions of incoming and outgoing waves. Thus they have roughly uniform energy for every shell of unit thickness. That leads to Eq.~(\ref{eq-rhoout}). Note that such an energy density would be divergent at $r=0$, but at scales shorter than the wavelength $n^{-1}$, it is smeared out and becomes uniform. That means the central ball of radius $n^{-1}$ has total energy equal to a large shell of thickness $n^{-1}$, which leads to Eq.~(\ref{eq-rhoin})\footnote{Throughout this paper, we use $\approx$ for an actual approximation, such as dropping subleading terms in large $n$. We use $\sim$ when we also drop all $n$-independent factors and are only interested in the scaling with $n$.}. In this physical picture, one can imagine a cutoff at $r=1$ and treat AdS as a finite box.

The enclosed mass in this single mode data can be estimated as
\begin{eqnarray}
M(r) = \int_0^r \rho_n(\bar{r})\bar{r}^{d-1}d\bar{r} 
&\sim& En^{d-1} r^d \ \ \ \ \ {\rm for} \ \ r\leq n^{-1}~, 
\label{eq-MassSingle} \\ \nonumber
&\sim& Er \ \ \ \ \ {\rm for} \ \  n^{-1}<r<1~.
\end{eqnarray}
Assuming the energy is dominated by the highest possible mode, $n\rightarrow\infty$, we can calculate the backreaction to the metric from.
\begin{eqnarray}
\delta g_{rr} \sim \frac{2M(r)}{r^{d-2}} &\sim& 2E r^{3-d}~.
\end{eqnarray}
Note that in $d=3$, this never diverges no matter how large $n$ is. On the other hand,
\begin{eqnarray}
\delta g_{tt} - \frac{2M}{r^{d-2}} = 4V(r) = -4\int_r^{1} (d-2)
\frac{M(\bar{r})}{\bar{r}^{d-1}}d\bar{r}
&\sim& 4E\ln r \ \ \ \ \ {\rm for} \ \ d=3~, 
\\ \nonumber
{\rm or} \ \ \ 
&\sim& -4E r^{3-d}  \ \ \ \ \ {\rm for} \ \ d>3~.
\end{eqnarray}
This diverges even for $d=3$.

Just from this simple example, we can see how ``energy cascade'' is a too na\"ive statement to conclude black hole formation. Even if all energy goes to one, infinitely high mode, it only means a black hole in $d>3$, since $\delta g_{rr}\sim-\delta g_{tt}$ indeed diverges together. In $d=3$, only $\delta g_{tt}$ diverges but $\delta g_{rr}$ does not. It is a large geometric backreaction, but not a black hole.

\subsection{Power law Spectrum}
\label{sec-PL}

Next we will consider a power law spectrum with 
\begin{equation}
A_n = A_0 (n+1)^{-\alpha}~.
\label{eq-PL}
\end{equation}
Even before any calculation, clearly, there should be some values of $\alpha$  large enough that the contribution from short wavelength modes is insufficient to make any feature in short distance scales. Likewise, there should be values of  $\alpha$ small enough that the short distance behavior is singular.  Our goal will be to find those thresholds.

First of all, we have to make a technical distinction between a finite power law, $\alpha>3/2$, and an infinite power law, $\alpha\leq3/2$. In the finite case, the spectrum has a finite IR amplitude for a finite total energy.
\begin{equation}
E_{\rm tot} =\sum_{n=0}^\infty w_n^2A_n^2 \sim \frac{A_0^2}{2\alpha - 3}~.
\end{equation}
In the infinite case, what we mean by an infinite sum is implicitly a limiting case when the UV cutoff $N$ on the sum goes to infinity, while the IR amplitude, $A_0$, drops to zero accordingly, maintaining a finite total energy.
\begin{eqnarray}
E_{\rm tot} = \sum_{n=0}^N w_n^2A_n^2   \label{eq-infpow}
&\sim& A_0^2 \ln N ~, \ \ \ {\rm for} \ \ \ \alpha = 3/2~, \\
&\sim& A_0^2 N^{3-2\alpha}~, \ \ \ {\rm for} \ \ \ \alpha<3/2~.
\end{eqnarray} 

In addition, there is a very important physical distinction when multiple eigenstates are involved---whether their phases, $B_n$ in Eq.~(\ref{eq-fourier}), are coherent or not. This directly plays a role in the calculation of mass enclosed.

\begin{eqnarray}
M(r) &\sim& \int_0^r r_1^{d-1}dr_1
\left( \sum_{n=0}^\infty w_n A_n e_n(r_1) \cos(w_nt+B_n) \right)^2~,
\label{eq-mass} \\ \nonumber
&\sim& \int_0^r r_1^{d-1}dr_1
\bigg[ \sum_{n=r^{-1}}^\infty w_n^2 A_n^2 e_n(r_1)^2 + 
\left(\sum_{n=0}^{r^{-1}} w_n A_n e_n(0) \cos(w_nt+B_n) \right)^2 \bigg]
\end{eqnarray}
For all modes with $n>r^{-1}$, they oscillate rapidly within the integration range, thus the cross terms automatically vanish from the integral. However, modes with $n<r^{-1}$ are basically constant within the integration range. If all of their phases are coherent, for example $t=\theta_n=0$ for all $n$, then the cross terms contribute significantly to the mass.

\subsubsection{Incoherent Phases}
\label{sec-incoherent}

Here we will derive the result when the phases are incoherent, thus all cross terms from Eq.~(\ref{eq-mass}) can be dropped.
\begin{eqnarray}
\label{eq-Mrandom}
M(r) &\sim& A_0^2 
\left( r \sum_{n=r^{-1}}^\infty n^{2-2\alpha} + r^d \sum_{n=0}^{r^{-1}} n^{d+1-2\alpha}\right)  
\\ \nonumber 
&\sim& E_{\rm tot} r^d \ \ \ \ \ \ \ \ \ \ \  {\rm for} \ \ \ \alpha>\frac{d+2}{2}~, \\ \nonumber
{\rm or} \ \ &\sim& -E_{\rm tot} r^d \ln r \ \ \ {\rm for} \ \ \ \alpha = \frac{d+2}{2}~,
\\ \nonumber
{\rm or} \ \ &\sim& E_{\rm tot} r^{2\alpha-2} \ \ \ \ \ \ \  {\rm for} 
\ \ \ \frac{3}{2}<\alpha<\frac{d+2}{2}~, 
\\ \nonumber
{\rm or} \ \ &\sim& E_{\rm tot} r \ \ \ \ \ \ \ \ \ \ \ \ {\rm for} \ \ \ \alpha\leq\frac{3}{2}~.
\end{eqnarray}
Note that we are only keeping the small $r$ behavior. Not surprisingly, for large $\alpha$, the behavior of lower modes dominates and the mass scales like volume, which is independent of $\alpha$. When the value of $\alpha$ drops below $(d+2)/2$, the energy density starts to develop a singularity at $r=0$, which grows more singular as $\alpha$ decreases further. Finally, infinite power laws lead to the same result as a single mode of arbitrarily high $n$, as seen in Eq.~(\ref{eq-MassSingle}).

A singular energy density implies a singular curvature tensor, but not always a large perturbation in the metric, which we will calculate here. For a more concise presentation, we only provide the explicit expression in the cases which the deviation can be singular. Any finite deviation will go to zero with $E_{\rm tot}$, so for our purpose their exact $r$ dependence is not that important.
\begin{eqnarray}
\delta g_{rr} \approx \frac{2M(r)}{r^{d-2}}
&\sim& 
{\rm regular}~, \ \ \ \ \ \ \ \ \ \ \ \ \ \ {\rm for} \ \ \ \alpha\geq\frac{d}{2}~, \\ \nonumber
{\rm or} \ \ &\sim& 
2E_{\rm tot} r^{2\alpha-d}~, \ \ \ \ \ \ \ \ \  \ \ {\rm for} \ \ \ \frac{3}{2}<\alpha<\frac{d}{2}~,
\ \ \ d>3~,
\\ \nonumber
{\rm or} \ \ &\sim&
2E_{\rm tot} r^{3-d}~, \ \ \ \ \ \ \ \ \ \ \ \ {\rm for} \ \ \ \alpha\leq\frac{3}{2}~.
\end{eqnarray}

\begin{eqnarray}
\delta g_{tt} - \frac{2M}{r^{d-2}} = 4V(r) &=& -4(d-2)\int_r^1\frac{M(r_1)}{r_1^{d-1}} dr_1 
\\ \nonumber
&\sim& {\rm regular}~, \ \ \ \ \ \ \ \ \ \ \ \ {\rm for} \ \ \ \alpha>\frac{d}{2}~, 
\\ \nonumber  {\rm or} \ \ \ &\sim& 
4E_{\rm tot} \ln r~, \ \ \ \ \ \ \ \ \ \ \ {\rm for} \ \ \ \alpha = \frac{d}{2}~, \ \ 
{\rm or} \ \ \alpha\leq\frac{3}{2}~, \ \ d=3~,
\\ \nonumber {\rm or} \ \ \ &\sim& 
-4E_{\rm tot} r^{2\alpha-d}~, \ \ \ \ \ \ \ {\rm for} \ \ \ \frac{3}{2}<\alpha<\frac{d}{2}~, 
\ \ d>3~,
\\ \nonumber {\rm or} \ \ \ &\sim&
-4E_{\rm tot} r^{3-d}~, \ \ \ \ \ \ \ \ {\rm for} \ \ \ \alpha\leq\frac{3}{2}~, \ \ d>3~.
\end{eqnarray}

As a quick summary, if the phases are incoherent, then approaching an $\alpha$-power law means
\begin{itemize}
\item Geometric deviation goes to zero with $E_{\rm tot}$ when $\alpha>d/2$, but curvature can already get large when $\alpha \leq (d+2)/2$. 
\item Geometric deviation gets large but {\bf does not} approach a black hole when $\alpha = d/2$ in any $d$ and $\alpha\leq3/2$ in $d=3$.
\item Geometric deviation gets large as if approaching a black hole when $\alpha<d/2$ in $d>3$, but it never does in $d=3$.
\end{itemize}

Just like in the single-mode case, $d=3$ is special. With incoherent phases, one never gets a black hole-like geometric deviation. Large geometric deviation still occurs when $\alpha\leq3/2$, namely for infinite power laws. That leads to $\delta g_{tt}$ blowing up like a log while $\delta g_{rr}$ stays small. That is also the situation in other dimensions with exactly $\alpha=d/2$. We can plug the behavior of of $\delta g_{rr}$ and $\delta g_{tt}$ into Eq.~(\ref{eq-Mink}) to get a better physical intuition for what is happening in these cases.
\begin{equation}
ds^2 = (1 - E_{\rm tot})\left(-r^{-E_{\rm tot}}dt^2 + dr^2\right) + r^2d\Omega_{d-1}^2~.
\end{equation}
There are order-one factors in front of both appearances of $E_{\rm tot}$ in the above equation that we did not keep track of, but those are not very relevant for our analysis. The point $r=0$ is singular for any positive $E_{\rm tot}$, but it takes only finite time for light rays to reach $r=0$ and come back to infinity, so it is not developing a horizon. It is a clear distinction between this deviation and those approaching a AdS-Schwarzschild metric.

\subsubsection{Coherent Phases}

In Sec.\ref{sec-lock} we will discuss more thoroughly what phase coherence means in this context.  Here let us just assume $t=\theta_n=0$ in Eq.~(\ref{eq-mass}). That leads to
\begin{eqnarray}
M(r) &\sim& A_0^2
\left[ r\sum_{n=r^{-1}}^\infty n^{2-2\alpha} + r^d\left(\sum_{n=0}^{r^{-1}} n^{\frac{d+1}{2}-\alpha} \right)^2 \right]~, 
\\ \nonumber
&\sim& E_{\rm tot} r^d~, \ \ \ \ \ \ \ \ \ \ \ {\rm for} \ \ \ \alpha>\frac{d+3}{2}~, 
\\ \nonumber {\rm or} \ \ \  &\sim&
E_{\rm tot}r^d(\ln r)^2~, \ \ \ {\rm for} \ \ \ \alpha = \frac{d+3}{2}~,
\\ \nonumber {\rm or} \ \ \ &\sim&
E_{\rm tot}r^{2\alpha-3}~, \ \ \ \ \ \ \ {\rm for} \ \ \ \frac{3}{2}<\alpha<\frac{d+3}{2}~.
\end{eqnarray}
We are omitting the technical results for infinite power laws here. Those cases have an ambiguity regarding the order of limits: the mode sum cut-off $N\rightarrow\infty$ and $r\rightarrow0$; they are also not too relevant for us since geometric deviation is already singular for finite power laws with small enough $\alpha$. Reducing $\alpha$ further to an infinite power law can only make the result more singular.

Note that when $\alpha\leq(d+3)/2$, the  energy density is already singular at $r=0$. As expected, this happens earlier (for a larger $\alpha$) compared to the case of incoherent phases in the previous session. It is straightforward to repeat the calculation of metric deviation and find that they also diverge earlier.
\begin{eqnarray}
\delta g_{rr} \approx \frac{2M(r)}{2r^{d-2}}
&\sim& 
{\rm regular}~, \ \ \ \ \ \ \ \ \ \ \ \ \ \ {\rm for} \ \ \ \alpha\geq\frac{d+1}{2}~, \\ \nonumber
{\rm or} \ \ &\sim& 
2E_{\rm tot} r^{2\alpha-d-1}~, \ \ \ \ \ \  \ \ {\rm for} \ \ \ 
\frac{3}{2}<\alpha<\frac{d+1}{2}~.
\end{eqnarray}
\begin{eqnarray}
\delta g_{tt} - \frac{2M}{r^{d-2}} = 4V(r) &=& -4(d-2)\int_r^1\frac{M(r_1)}{r_1^{d-1}} dr_1 
\\ \nonumber
&\sim& {\rm regular}~, \ \ \ \ \ \ \ \ \ \ \ \ {\rm for} \ \ \ \alpha>\frac{d+1}{2}~, 
\\ \nonumber  {\rm or} \ \ \ &\sim& 
4E_{\rm tot} \ln r~, \ \ \ \ \ \ \ \ \ \ \ {\rm for} \ \ \ \alpha = \frac{d+1}{2}~, 
\\ \nonumber {\rm or} \ \ \ &\sim& 
-4E_{\rm tot} r^{2\alpha-d-1}~, \ \ \ \ {\rm for} \ \ \ \frac{3}{2}<\alpha<\frac{d+1}{2}~.
\end{eqnarray}
We have collected all of these results in the summary table at the beginning of this section.

\section{Stationary Coherent Power Law Solutions}
\label{sec-lock}
In this section we examine the evolution equations, using the two-time formalism, to establish the existence of phase-coherent power laws as exactly stationary solutions. 

\subsection{Two-Time Analysis}
\label{sec-ttf}

%In the recent literature about the AdS (in)stability problem, one often starts from the following coordinate system of empty $AdS_{d+1}$,
%\begin{equation}
%ds^2 = \frac{1}{\cos^2x}
%\left(-dt^2 + dx^2 + \sin^2x~d\Omega_{d-1}^2\right)~,
%\end{equation}
We first review the two-time formalism that is employed to describe the AdS-gravity dynamics at the $\epsilon^{-2}$ time scale \cite{BalBuc14,CraEvn14a,BizMal15}. A spherically symmetric free scalar field in a fixed AdS background can be decomposed into eigenstates\cite{HamKab06,HamKab06a},
\begin{eqnarray}
\phi(t,r) &=& \sum_{n=0}^\infty \phi_n(t)e_n(r) \equiv \sum_{n=0}^\infty \bar{A}_n e_n(r) \cos(w_nt + B_n)~, 
\label{eq-fourier} \\
e_n(r) &=& \sqrt{\frac{(2n+d)n!\Gamma(n+d)}
{2^d\Gamma(n+d/2)\Gamma(n+d/2+1)}}
(1+r^2)^{-d/2} ~ P_n^{(d/2-1,d/2)}\left(\frac{1-r^2}{1+r^2}\right)~.
\label{eq-eigen}
\end{eqnarray}
The eigenfrequencies are all integers given by $w_n = 2n + d$. 

Without gravity, $\bar{A}_n$ and $B_n$ will stay constant forever. Including gravity, the presence of energy from this field modifies the metric, which in turn modifies the evolution of the field. When such effect is small, it can be approximated by\footnote{The constraint $k+l=m+n$ is the combination of two effects. (1) The resonant condition $w_n = \pm w_k \pm w_l \pm w_m$, and (2) the actual evaluation of $C_{klmn}$ which is related to hidden symmetries of $AdS$ and extra conserved quantities in the dynamics \cite{Yan15,EvnKri15,EvnNiv15}.}
\begin{equation}
\ddot\phi_n + w_n^2\phi_n = \sum_{k,l,m}^{k+l=m+n} C_{klmn} \phi_k\phi_l\phi_m 
+ \mathcal{O}(\phi^5)~. 
\label{eq-eom}
\end{equation}
The stability at the $T\sim\epsilon^{-2}$ time scale, taking the $\epsilon\rightarrow0$ limit, can always be addressed within the regime that the higher order terms can be safely dropped \cite{DimYan15}. This is effectively a collection of quartically coupled harmonic oscillators. 

One can rewrite this second order differential equation into two first order equations for  $\bar{A}_n$ and $B_n$.
\begin{eqnarray}
2w_n\frac{d\bar{A}_n}{dt} &=& \sum_{klm}^{k+l=m+n}  
C_{klmn} \bar{A}_k \bar{A}_l \bar{A}_m 
\sin (B_n+B_m-B_k-B_l)~, \\
2w_n\frac{dB_n}{dt} &=& \bar{A}_n^{-1} \sum_{klm}^{k+l=m+n}  
C_{klmn}\bar{A}_k \bar{A}_l \bar{A}_m 
\cos (B_n+B_m-B_k-B_l)~.
\end{eqnarray}
Note that we can rescale time, $t = \tau\epsilon^{-2}$, and also rescale the amplitudes, $\bar{A}_n = A_n\epsilon$, we can the dynamical equations in the ``long time'' $\tau$.
\begin{eqnarray}
2w_n\frac{dA_n}{d\tau} &=& \sum_{klm}^{k+l=m+n}  C_{klmn} A_k A_l A_m 
\sin (B_n+B_m-B_k-B_l)~, 
\label{eq-2tfA} \\
2w_n\frac{dB_n}{d\tau} &=& A_n^{-1} \sum_{klm}^{k+l=m+n}  C_{klmn}A_k A_l A_m 
\cos (B_n+B_m-B_k-B_l)~.
\label{eq-2tfB}
\end{eqnarray}
This set of equations then represents the evolution of the scale-independent, relative amplitudes of all modes, together with their phases.

Note that in the previous section, we have chosen a gauge that the time at the asymptotic boundary stays the same, thus our equations here are also in such boundary gauge. As discussed in \cite{CraEvn14}, such gauge is intuitively convenient since there exists a Lagrangian (and Hamiltonian) that reproduces the equations of motion. Furthermore, the point $r=0$ is quite special in the spherically symmetric setup, and using its proper time can be misleading. For example, the oscillating divergence observed in \cite{BizMal15} means that the point $r=0$ has an infinite redshift with respect to any other point. Whether it means a black hole is unclear, as we already explained in Sec.\ref{sec-kinematics}. In other to avoid similar confusions, in the rest of this paper we will stay in this boundary gauge unless otherwise specified.

Physically, the phases are coherent if there is some time during one AdS period where all of the modes are in phase. The phase $\theta_n$ of the mode $n$ is related to the ``slow phase" $B_n$ by
\begin{equation}
\theta_n (\tau, t) = B_n(\tau) + \omega_n t = B_n(\tau) + (2 n + d) t
\end{equation}
Note that the slow phase $B_n$ depends on the slow time $\tau$, while the full phase $\theta_n$ depends on the fast time as well. 

For the phases to align at some time during the short time period $\delta t = 2 \pi$ requires
\begin{equation}
\theta_n(\tau, t) - \theta_m (\tau, t) =  2 \pi N_{nm}~,
\end{equation}
where $N_{nm}$ are integers that can depend on the modes involved. Coherence requires that we can solve this equation for the short time $t$ over one cycle $0< t < 2 \pi$, at the same $t$ for all modes. Plugging in the formula for the phases $\theta_n$, we have
\begin{equation}
B_n(\tau) - B_m(\tau) = 2 \pi N_{nm} + 2 (n-m) t
\end{equation}
Since the $B_n$ are only defined mod $2 \pi$, we can drop the first term on the right side. Define $2 t \equiv \theta(\tau)$, the equation becomes simply
\begin{equation}
B_n(\tau) - B_m(\tau) = (n - m) \theta(\tau)~.
\end{equation}
Solving this equation for all choices of $m$ and $n$ requires
\begin{equation}
B_n(\tau) =  n \gamma(\tau) + \delta(\tau)~,
\label{eq-phaseco}
\end{equation}
where $\gamma, \delta$ are free functions of the slow time that must be independent of the mode number $n$, and the equation is valid mod $2 \pi$.

We are interested in describing the behavior at large mode numbers, so we should allow corrections to this formula. Our final condition for phase coherence is therefore
\begin{equation}
B_n(\tau) =  n \gamma(\tau) + \delta(\tau) + ...
\label{eq-coherent}
\end{equation}
Here ``...'' are just anything that goes to zero in the large $n$ limit. It may be interesting to consider a weaker notion of phase coherence, which would still allow for constructive interference in the gravitational backreaction, but in this paper we will only use the above definition.

\subsection{Asymptotic Phase-Coherent Power Laws}
%\subsubsection{Higher dimensions}
%\label{sec-higher}
We now want to self-consistently solve the slow-time evolution equations, Eq.~(\ref{eq-2tfA}) and (\ref{eq-2tfB}), under the coherent phase condition Eq.~(\ref{eq-coherent}). In order to analyze the equations, we need to know the scaling of the interaction coefficients $C_{ijkl}$. In \cite{CraEvn15a}, it was reported that in the boundary gauge, the coefficients obey the simple scaling law,
\begin{equation}
C_{(\lambda k)(\lambda l)(\lambda m)(\lambda n)} \sim \lambda^{d}
C_{klmn}~,
\label{eq-Cscale}
\end{equation}
for greater than 3 spatial dimensions, $d >3$.

In a forthcoming publication \cite{DFPY15}, we find that in fact this scaling is modified for the diagonal terms $C_{iijj}$ and $C_{iiii}$ in $d=4$ by additional logarithmic factors; however these factors do not appear to affect the final results, so here we use the simple scaling in Eq.~(\ref{eq-Cscale}) and defer a more detailed description to \cite{DFPY15}. In higher dimensions, $d > 4$, the scaling (\ref{eq-Cscale}) is exact for large mode numbers.
% In the central gauge, which has been much used previously, the diagonal terms scale differently from the off-diagonal terms. This complicates the arguments below, so we do the analysis in the boundary gauge.

First of all, the phase-locked condition, Eq.~(\ref{eq-coherent}), is already a natural solution to one of the equations of motion, Eq.~(\ref{eq-2tfA}). Since the resonant condition is $m+n=k+l$, the phase-locked condition makes $(B_n+B_m-B_k-B_l)=0$. This makes all the sine terms in Eq.~(\ref{eq-2tfA}) zero, thus $dA_n/d\tau=0$. In other words, this choice of phases guarantees that there is no energy transfer among the modes. This is exactly the same as in the quasi-periodic, non-collapsing solutions \cite{GreMai15}.  The remaining question is whether the coherent phase assumption is maintained under time evolution.

Examining the equation for the phase evolution (\ref{eq-2tfB}), all the cosine factors there are 1 due to the coherent phase ansatz, so this equation takes a very simple form.
\begin{eqnarray}
2w_n\frac{dB_n}{d\tau}
&=& A_n^{-1} \sum_{k,l,m}^{k+l=m+n}  C_{klmn}A_k A_l A_m~.
\label{eq-BC}
\end{eqnarray}

We need $B_n = n \gamma(\tau) + \delta(\tau) + ...$ to maintain the phase coherence, and $\omega_n \sim n$, so the left side of the equation must have the n-scaling $n^2  \gamma(\tau) + n \delta(\tau)$. As long as $\gamma(\tau) \neq 0$, this 
means that the right-hand-side of the above equation must scale like $n^2$. Plugging in the power law spectrum, $A_n = A_0 n^{- \alpha}$, into the right side of Eq.~(\ref{eq-BC}), we get
\begin{equation}
2w_n\frac{dB_n}{d\tau} =
A_0^2 n^\alpha \sum_{k,l,n}^{k+l=m+n}  C_{klmn} [k(m+n-k)m]^{-\alpha}~.
\end{equation}
Then we use integrals to approximate the sums.
\begin{equation}
\approx A_0^2 n^{\alpha} \int dk~dm~C_{k(m+n-k)mn}~[k(m+n-k)m]^{-\alpha} ~.
\label{eq-intapp}
\end{equation}
Whether the integral approximation to the sums is a good one depends on the detailed dependence of the coefficients $C_{ijkl}$ on each one of the indices, not only on the overall scaling. For now, we assume the integral approximation holds, and leave a more careful analysis for future work. 
%In order for this approximation to work, we need the values of $C_{klmn}$ to behave regularly enough to have a smooth continuum limit. That is indeed the case and for physical quantities it is often true. In addition, we require that the $k$ integral does not diverge at the integration boundaries. Note that the original sum of $k$ has only finite terms, thus never diverge anyway. Divergence in the integral approximation simply means that the sum is dominated by a fixed few terms near $k=0$ and $k=(m+n)$, thus it will not scale like a sum correctly. That could be checked by the condition
%\begin{equation}
%C_{1(m+n-1)mn} A_1A_{m+n-1}\lesssim (m+n) 
%C_{\frac{m+n}{2}\frac{m+n}{2}mn} A_{\frac{m+n}{2}}^2~,
%\label{eq-intcond}
%\end{equation}
%for large $(m+n)$.

Within the integral approximation, we can utilize the scaling behavior in Eq.~(\ref{eq-Cscale}).
\begin{eqnarray}
& &A_0^2 n^{\alpha} \int dk~dm~C_{k(m+n-k)mn}~[k(m+n-k)m]^{-\alpha} \\
&=& A_0^2 n^{\alpha} \int n^2 dx ~ dz ~
n^d C_{x(z+1-x)z1}~n^{-3\alpha}[x(z+1-x)z]^{-\alpha}  \\ \nonumber
&\propto& A_0^2~n^{d+2-2\alpha}~.
\label{eq-scaling}
\end{eqnarray}
Thus $\alpha = d/2 $ is the unique value to provide $n^2$ scaling, maintaining the phase-locked condition. If we had considered the special case $\gamma = 0$, then the self-consistent solution would be a different power law, $\alpha = d/2 + 1/2$. In the doubly special case $\gamma = \delta =0$, the value is $\alpha = d/2 + 1$. By examining the early time dynamics in the Section \ref{sec-seed}, we believe that the generic case $\gamma \neq 0$ is dynamically selected.

Note that $\alpha=d/2$ we find here, strictly speaking, is necessary but not sufficient for the phases to remain coherent dynamically. It forbids higher order $n$ scaling in $B_n$, but it is not clear whether there are subleading fractional powers of $n$ or order 1 fluctuating contributions. Those can potentially ruin the phase coherence, but could only be checked given subleading behavior of the coupling coefficients $C_{ijkl}$. These are difficult to obtain. 

Leaving these various caveats aside, we can go ahead and ask whether the power law predicted by our analysis agrees with that observed in the full numerical evolution. Maliborski and Rostworowski \cite{MalRos13} suggested a ``preliminary guess" for the energy spectrum
\begin{equation}
E_n \sim n^{- {6 \over 5} - {4 \over 5}(d-3)}~.
\end{equation}
The energy per mode is related to the amplitude by $E_n \sim \omega_n^2 A_n^2 \sim n^{2 - 2 \alpha}$. Plugging in our values of $\alpha$,  our analysis predicts an energy spectrum
\begin{equation}
E_n \sim n^{2 - d}~.
\end{equation}

Recall that we have assumed the scaling (\ref{eq-Cscale}), which is valid in $d>4$, and almost valid (up to logarithmic corrections) in $d=4$. In $d=4$, our formula agrees with the Maliborski-Rostworowski guess. In $d=5$, we get $E_n \sim n^{-3}$, while the M-R formula gives $n^{-2.8}$. It is not a big difference, and there has not been a lot of data to accurately determine the actual power yet\footnote{We thank A. Rostworowski for sharing his results in private communications.}. Further numerical results in $d\geq5$ would provide an important check for our predictions.

\subsubsection{$AdS_{3+1}$}

The situation in $AdS_{3+1}$ is trickier. 
%First of all, the observed power law, $\alpha=6/5$ is a diverging power law. It requires some special treatment as we discuss Appendix \ref{sec-PL}. In order to maintain finite total energy, we have to first put a UV cut-off on the sum, and simultaneously take that to infinity as the IR amplitude, $A_0$, goes to zero.
As we will explain in \cite{DFPY15}, the diagonal terms in the boundary gauge are still likely to have peculiar behaviors, thus such a gauge choice does not simplify the matter here. This subsection will be an exception to the rest of the paper, and we will use the central gauge, in which the scaling property of $C_{klmn}$ was analytically derived in \cite{GreMai15,CraEvn15b}.
\begin{equation}
C_{jnjn} \sim n^2 j^2 \ln j~, \ \ \ \ \ \ 
C_{(\lambda k) (\lambda l) (\lambda m) (\lambda n)} \sim \lambda^3 C_{klmn}~.
\end{equation}
Fortunately, the diagonal terms trivially satisfy the phase-locked requirement due to its $n^2$ dependence\footnote{Even though the sum over diagonal terms may appear to be logarithmically diverging, indicating some mild cutoff dependence, the $n^2$ factor still guarantees that it does not ruin phase coherence. This apparent divergence is mitigated or eliminated in the boundary gauge.}. The remaining question is the off-diagonal terms, which are the same in either gauge.

These terms can then be analyzed in exactly the same way as above. Naively extending our result to $d=3$ gives $A_n \sim n^{-3/2}$, or equivalently $E_n \sim n^{-1}$. Note that an energy spectrum $n^{-1}$ is not normalizable at large $n$, so we need to do a more refined analysis as explained in Sec.\ref{sec-PL}: Including a UV cutoff $N$ such that the amplitudes go to zero as $N$ goes to infinity to conserve total energy. Keeping that in mind, we can begin with a  similar process.

% \begin{equation}
% E_n = {E_{\rm tot} \over \log N} n^{-1} \ \ \ \ \ \ \  A_n = \sqrt{E_{\rm tot} \over \log N} n^{-3/2}
% \end{equation}
% where $E_{\rm tot}$ is the total energy.

% Plugging this into (\ref{eq-scaling}), we find the requirement
% \begin{equation}
% A_0^2 n^2 = {E \over \log N} n^2 \sim n^2
% \end{equation}
% which is satisfied. 

\begin{eqnarray}
2w_n\frac{dB_n}{d\tau}
&=& A_n^{-1} \sum_{k,l,m}^{k+l=m+n}  C_{klmn}A_k A_l A_m~, \\ \nonumber
&\approx& A_0^2 n^{\alpha} \int_0^N dm~\int_0^{m+n} dk~
C_{k(m+n-k)mn}~[k(m+n-k)m]^{-\alpha}~,
\\ \nonumber 
&\approx& A_0^2 n^{5-2\alpha} \int_0^{\frac{N}{n}}dy~\int_0^{y+1}dx~
C_{x(y+1-x)y1}[x(y+1-x)y]^{-\alpha}
\label{eq-AdS3-1}
\end{eqnarray}
Here $N$ is the UV cut-off which will later go to infinity as $A_0$ goes to zero according to Eq.~(\ref{eq-infpow}). After scaling out $n$, in the rescaled integral, only at most 3 out of 4 indices will get large, and those particular coefficients scale quite differently. Such scaling behavior was derived in \cite{CraEvn15b}.
\begin{eqnarray}
C_{j(\lambda m +\lambda n)(\lambda m)(\lambda n)} \sim \lambda C_{j(m+n)mn}~
&,& \ \ \ {\rm with} \ j\ll m,n~, \\
C_{k(\lambda l)(\lambda m)n} \sim \lambda^2 C_{klmn}~
&,& \ \ \ {\rm with} \ l,m\gg k,n~.
\end{eqnarray}
Note that the coefficients with 2 large indices are actually 1 power of $\lambda$ higher than those with large 3 indices. This means that one can use either the $\lambda$ scaling to analyze the double integral, or simply keep the boundary terms of the $x$ integral and use the $\lambda^2$ scaling. They lead to the same answer.
\begin{eqnarray}
& & A_0^2 n^{5-2\alpha} \int_0^{\frac{N}{n}}dy~\int_0^{y+1}dx~
2C_{x(y+1-x)y1}[x(y+1-x)y]^{-\alpha} \\ \nonumber
&\approx& A_0^2 n^{5-2\alpha}  \int_0^{\frac{N}{n}}dy~
2C_{0(y+1)y1}[(y+1)y]^{-\alpha}~, \\ \nonumber
&\approx& A_0^2 n^{5-2\alpha} \left(\frac{N}{n}\right)^{3-2\alpha}
\sim E_{\rm tot} n^2~.
\end{eqnarray}
The fact that the $A_0^2 N^{3-2\alpha}$ combination correctly reduces to the finite total energy is a good assurance that our estimation is reasonable. The case with $\alpha=3/2$ will produce a log in the second last step but also cancels out exactly to reach the the same final answer.

Quite interestingly, the $n^2$ scaling, thus the phase-lock condition, is guaranteed by any divergent power law, thus provides an upper bound $\alpha\leq3/2$. Extensive numerics has been done in 3+1 dimensions, and the most up-to-date result seems to suggest $E_n \sim n^{- 6/5}$ \cite{BizMal15}, namely $\alpha = 8/5$, which slightly exceeds our upper bound. Note that our bound requires an infinite power law, and any actual numerical study must have a UV cut-off. It is possible that such cut-off forbids the power law to be exactly achieved. In the future, one can try to check whether pushing to higher cut-off makes the value of $\alpha$ closer to $3/2$. If the current value of $\alpha = 6/5$ is confirmed, then one of our assumptions must be wrong. One obvious candidate is that it may be wrong to replace the sums by integrals.

\section{Initial Phase Coherence}
\label{sec-seed}

Note that the phase-coherent solutions are not guaranteed to be attractors. Even if the phase $B_n(\tau)$ is dominated by a term proportional to $n$ at late times, we cannot just drop the subleading terms. The phases only matter mod $2\pi$, thus any finite contribution matters. In fact, even the initial phases are relevant throughout the entire process. Here we will demonstrate that the two-mode initial data, an initial condition that has been frequently tested to lead to collapse, provide an appropriate initial condition leading to coherent phases.

The two-mode initial data is given by $A_0 \sim A_1 \sim \epsilon$ with arbitrary initial phases $B_0$ and $B_1$. For $t\ll\epsilon^2$, namely $\tau\ll1$, we can pretend that $\phi_0$ and $\phi_1$ stay as the free eigenstates, and solve higher modes in Eq.~(\ref{eq-eom}) as being resonantly driven, starting from zero amplitudes, by the lower ones. For example, $\phi_2$ obeys the equation
\begin{eqnarray}
\ddot\phi_2 + w_2^2\phi_2 = S_{1102}\phi_1^2\phi_0 
\sim \epsilon^3 \cos\left[(2w_1-w_0)t + (2B_1-B_0)\right]~.
\end{eqnarray}
where in the last equality we have only kept the source terms that are in resonance. This is solved by
\begin{equation}
\phi_2 \sim \epsilon^3 t \cos\left[(2w_1-w_0)t + (2B_1-B_0)-\pi/2\right]~.
\label{eq-mode2}
\end{equation}
Again we have dropped some order-one factors. We only care about the powers of $\epsilon$ and $t$, and the phases. The above behavior for $\phi_2$ is nothing but the well-known fact that a constant amplitude, resonant driving force will lead to a linear growth. It is actually a special case of a ``polynomially driven'' harmonic oscillator,
\begin{equation}
\ddot{f} + w^2f = C t^j\cos(wt+\theta_j)~,
\label{eq-PolyDrive}
\end{equation}
with the solution
\begin{equation}
f \sim t^{j+1}\cos(wt+\theta_j-\pi/2)~, 
\label{eq-PolySolve}
\end{equation}
as we will show in Appendix \ref{sec-PDHO}.

Using this general polynomial growth, one can show that higher modes, during the time $1\ll t \ll \epsilon^{-2}$, are given by the following general form. 
\begin{eqnarray}
\phi_n \sim \epsilon \left(\epsilon^2t\right)^{n-1}
\cos\left[w_nt +(n-1)(B_1-B_0-\pi/2)+B_1\right]
\label{eq-moden}
\end{eqnarray}
To establish this, note that 
Eq.~(\ref{eq-mode2}) is not only the special case with $n=2$, but also the first step for a proof of mathematical induction. The next step is to assume that Eq.~(\ref{eq-moden}) is true for all $2\leq i <n$, and show that it holds for $\phi_{n+1}$. We can show this as follows:
\begin{eqnarray}
\ddot\phi_n + w_n\phi_n &=& \sum_{0\leq k,l,m<n}^{k+l = m+n} C_{klmn}\phi_k\phi_l\phi_m 
\label{eq-TreeTurtle}
\\ \nonumber
&\sim& \sum_{1\leq k,l,m<n}^{k+l = m+n}
\epsilon^3 \left(\epsilon^2 t\right)^{k+l+m-3} 
\cos\left[w_nt + (k+l-m-1)(B_1-B_0-\pi/2) +B_1 \right] \\ \nonumber
& & + \sum_{1\leq k,l<n}^{k+l=n}
\epsilon^3 \left(\epsilon^2 t\right)^{k+l-2} 
\cos\left[w_nt + (k+l-2)(B_1-B_0-\pi/2) +2B_1-B_0 \right] 
\\ \nonumber
&\sim& \epsilon^3\left(\epsilon^2 t\right)^{n-2}
\cos [w_n t + (n-2)(B_1-B_0-\pi/2 )+ 2 B_1-B_0]
\end{eqnarray}
The key point allowing for the simplification is that the sum is dominated by terms with $m=0$, since it has the lowest power of $\epsilon$. Note that  these terms have the same phase, which will be true as long as the initial amplitudes of two modes are comparable and dominate over others. Thus, the last line in 
Eq.~(\ref{eq-TreeTurtle}) has only one phase just like Eq.~(\ref{eq-PolyDrive}), with 
\begin{equation}
\theta_n = (n-2)(B_1-B_0-\pi/2) + 2 B_1-B_0 ~.
\end{equation}
Thus the solution $\phi_n$ is given by Eq.~(\ref{eq-PolySolve}), which indeed proves Eq.~(\ref{eq-moden}). We can now identify the phases $B_n$ in this regime,
\begin{equation}
B_n = \theta_n - \pi/2 = n(B_1 - B_0 - \pi/2) + B_0 + \pi/2
\end{equation}
These phases are coherent in the sense of Eq.~(\ref{eq-phaseco}). Furthermore, since in the early stage the phases $B_n$ already develop a linear $n$ dependence, we think it is natural for the late time asymptotics to maintain such behavior, thus we should focus on the $\gamma\neq0$ case in Eq.~(\ref{eq-coherent}).

Note that every dominant term having the same phase in Eq.~(\ref{eq-TreeTurtle}), independent of the initial amplitudes (as long as it is two-mode dominated), is a very special property. A three-mode initial data would have immediately undermined our simple analysis. Thus we can see that the two-mode initial data is particularly appropriate to provide initially coherent phases. It would be very interesting to extend this type of analysis to more general initial data. This could give insight into which initial data evolve into a coherent cascade.

\acknowledgments

We thank Alex Buchel, Stephen Green, Fotios Dimitrakopoulos, Luis Lehner, Matt Lippert, Juan Pedraza, and Andrzej Rostworowski, and the anonymous referee from Physical Review D for useful discussions. B.F. is part of the $\Delta$-ITP consortium and also supported in part by the Foundation for Fundamental Research on Matter (FOM), both are parts of the Netherlands Organisation for Scientific Research (NWO) that is funded by the Dutch Ministry of Education, Culture and Science (OCW). I.S.Y. is supported by the Government of Canada through Industry Canada and by the Province of Ontario through the Ministry of Research and Innovation.

\appendix

\section{Random Phase Ansatz}
\label{sec-random}

We can rewrite the two-time equation of motion for the amplitude $A_n$ and phase $B_n$, Eq.~(\ref{eq-2tfA}) and (\ref{eq-2tfB}), as one complex equation for $a_n = A_ne^{iB_n}$.
\begin{equation}
2iw_n\dot{a}_n = \sum_{klm}C_{klmn}a_ka_la_m^*~.
\label{eq-tteom}
\end{equation}
Here dot is the derivative with respect to the ``long time'' $\tau$. The random phase ansatz assumes that the phases $B_n$ is randomly distributed between $0$ and $2\pi$, with a constant weight, and every mode is independent from one another. Using this statistics property, we know the property of any two-mode correlator while averaging over an ensemble of random phases.
\begin{equation}
\langle a_m a_n \rangle=0~, \ \ \ 
\langle a_m a_n^*\rangle = \frac{N_n}{w_n}\delta_{mn}~. \ \ \ 
\end{equation}
Here $N_n$ is the expectation value of ``particle number'' in a mode as defined in \cite{CraEvn14a}. We also know the behavior of any four-mode correlator since it factorizes.
\begin{equation}
\langle a_k a_l a_m^* a_n^* \rangle = N_m N_n 
\left(\delta_{km}\delta_{ln} + \delta_{kn}\delta_{lm} \right)~.
\label{eq-4pt}
\end{equation}
It is then easy to show that
\begin{eqnarray}
\dot{N}_n &=& w_n
\left( \langle \dot{a}_n a_n^* \rangle + \langle a_n\dot{a}_n^* \rangle \right)
\\ \nonumber
&=& \frac{w_n}{2}
\left(-i\sum_{klm}C_{klmn}\langle a_ka_la_m^*a_n^*\rangle + i\sum_{klm}C_{klmn}\langle a_k^*a_l^*a_ma_n \rangle \right)=0~.
\end{eqnarray}
In the last step, we simply plug in Eq.~(\ref{eq-4pt}).

This proves that if we combine two-time formalism with the random phase ansatz, we will get no dynamics at the leading order time scale of the two-time formalism.

\section{Polynomially Driven Harmonic Oscillator}
\label{sec-PDHO}

In the main text we needed the solution to a ``polynomially driven" oscillator, satisfying the equation
\begin{equation}
\ddot{f} + w^2f = C t^j\cos(wt+\theta_j)~.
\end{equation}

First we assume that the solution is
\begin{eqnarray}
f = \sum_{i=0}^{j+1} c_i t^i\cos(wt+\xi_i)~.
\end{eqnarray}
Taking derivatives and rearranging the sum, we get
\begin{eqnarray}
\ddot{f} + w^2 f &=& -2(j+1) t^n c_{j+1} w \sin(wt+\xi_{i+1}) \\ \nonumber
&+& \sum_{i=1}^j t^{i-1} 
\left[-2ic_iw\sin(wt+\xi_i) + i(i+1)c_{i+1}\cos(wt+\xi_{i+1})\right]~.
\end{eqnarray}

This can be solved recursively as
\begin{eqnarray}
c_{j+1} &=& \frac{C}{2w(j+1)}~, \\
c_i &=& \frac{(j+1)!c_{j+1}}{(2w)^{j+1}}\frac{(2w)^i}{i!}~, \\
\xi_{j+1} &=& \theta_j - \pi/2~, \\
\xi_{i} &=& \xi_{i+1}+\pi/2~.
\end{eqnarray}

Whenever $(wt)\gg1$, the solution is dominated by the highest polynomial,
\begin{equation}
f(t) \approx \frac{C}{2w}\frac{t^{j+1}}{j+1} \cos(wt+\theta_j-\pi/2)~.
\end{equation}

\bibliographystyle{utcaps}
\bibliography{all}

\providecommand{\href}[2]{#2}\begingroup\raggedright\begin{thebibliography}{10}

\bibitem{ChrKla93}
D.~Christodoulou and S.~Klainerman, ``The global nonlinear stability of the
  Minkowski space,'' {\em S{\'e}minaire {\'E}quations aux d{\'e}riv{\'e}es
  partielles (Polytechnique)} (1993)  1--29.

\bibitem{Fri86}
H.~Friedrich, ``Existence and structure of past asymptotically simple solutions
  of Einstein's field equations with positive cosmological constant,'' {\em
  Journal of Geometry and Physics} {\bf 3} (1986) no.~1, 101--117.

\bibitem{BizRos11}
P.~Bizon and A.~Rostworowski, ``{On weakly turbulent instability of anti-de
  Sitter space},'' \href{http://dx.doi.org/10.1103/PhysRevLett.107.031102}{{\em
  Phys.Rev.Lett.} {\bf 107} (2011)  031102},
\href{http://arxiv.org/abs/1104.3702}{{\tt arXiv:1104.3702 [gr-qc]}}.
%%CITATION = ARXIV:1104.3702;%%.

\bibitem{DiaHor11}
O.~J. Dias, G.~T. Horowitz, and J.~E. Santos, ``{Gravitational Turbulent
  Instability of Anti-de Sitter Space},''
  \href{http://dx.doi.org/10.1088/0264-9381/29/19/194002}{{\em
  Class.Quant.Grav.} {\bf 29} (2012)  194002},
\href{http://arxiv.org/abs/1109.1825}{{\tt arXiv:1109.1825 [hep-th]}}.
%%CITATION = ARXIV:1109.1825;%%.

\bibitem{Lie12}
S.~L. Liebling, ``{Nonlinear collapse in the semilinear wave equation in AdS
  space},'' \href{http://dx.doi.org/10.1103/PhysRevD.87.081501}{{\em Phys.Rev.}
  {\bf D87} (2013) no.~8, 081501},
\href{http://arxiv.org/abs/1212.6970}{{\tt arXiv:1212.6970 [gr-qc]}}.
%%CITATION = ARXIV:1212.6970;%%.

\bibitem{deOPan12}
H.~de~Oliveira, L.~A. Pando~Zayas, and E.~Rodrigues, ``{A Kolmogorov-Zakharov
  Spectrum in AdS Gravitational Collapse},''
  \href{http://dx.doi.org/10.1103/PhysRevLett.111.051101}{{\em Phys.Rev.Lett.}
  {\bf 111} (2013) no.~5, 051101},
\href{http://arxiv.org/abs/1209.2369}{{\tt arXiv:1209.2369 [hep-th]}}.
%%CITATION = ARXIV:1209.2369;%%.

\bibitem{DiaHor12}
O.~J. Dias, G.~T. Horowitz, D.~Marolf, and J.~E. Santos, ``{On the Nonlinear
  Stability of Asymptotically Anti-de Sitter Solutions},''
  \href{http://dx.doi.org/10.1088/0264-9381/29/23/235019}{{\em
  Class.Quant.Grav.} {\bf 29} (2012)  235019},
\href{http://arxiv.org/abs/1208.5772}{{\tt arXiv:1208.5772 [gr-qc]}}.
%%CITATION = ARXIV:1208.5772;%%.

\bibitem{BucLie13}
A.~Buchel, S.~L. Liebling, and L.~Lehner, ``{Boson stars in AdS spacetime},''
  \href{http://dx.doi.org/10.1103/PhysRevD.87.123006}{{\em Phys.Rev.} {\bf D87}
  (2013) no.~12, 123006},
\href{http://arxiv.org/abs/1304.4166}{{\tt arXiv:1304.4166 [gr-qc]}}.
%%CITATION = ARXIV:1304.4166;%%.

\bibitem{MalRos13}
M.~Maliborski and A.~Rostworowski, ``{Lecture Notes on Turbulent Instability of
  Anti-de Sitter Spacetime},''
  \href{http://dx.doi.org/10.1142/S0217751X13400204}{{\em Int.J.Mod.Phys.} {\bf
  A28} (2013)  1340020},
\href{http://arxiv.org/abs/1308.1235}{{\tt arXiv:1308.1235 [gr-qc]}}.
%%CITATION = ARXIV:1308.1235;%%.

\bibitem{MalRos13a}
M.~Maliborski and A.~Rostworowski, ``{Time-Periodic Solutions in an Einstein
  AdS–Massless-Scalar-Field System},''
  \href{http://dx.doi.org/10.1103/PhysRevLett.111.051102}{{\em Phys.Rev.Lett.}
  {\bf 111} (2013) no.~5, 051102},
\href{http://arxiv.org/abs/1303.3186}{{\tt arXiv:1303.3186 [gr-qc]}}.
%%CITATION = ARXIV:1303.3186;%%.

\bibitem{AllFir14}
A.~Allahyari, J.~T. Firouzjaee, and R.~Mansouri, ``{Gravitational collapse in
  the AdS background and the black hole formation},''
\href{http://arxiv.org/abs/1404.7783}{{\tt arXiv:1404.7783 [gr-qc]}}.
%%CITATION = ARXIV:1404.7783;%%.

\bibitem{MalRos14}
M.~Maliborski and A.~Rostworowski, ``{What drives AdS unstable?},''
\href{http://arxiv.org/abs/1403.5434}{{\tt arXiv:1403.5434 [gr-qc]}}.
%%CITATION = ARXIV:1403.5434;%%.

\bibitem{BalBuc14}
V.~Balasubramanian, A.~Buchel, S.~R. Green, L.~Lehner, and S.~L. Liebling,
  ``{Holographic Thermalization, Stability of AdS, and the FPU Paradox},''
\href{http://arxiv.org/abs/1403.6471}{{\tt arXiv:1403.6471 [hep-th]}}.
%%CITATION = ARXIV:1403.6471;%%.

\bibitem{OkaCar14}
H.~Okawa, V.~Cardoso, and P.~Pani, ``{On the nonlinear instability of confined
  geometries},''
\href{http://arxiv.org/abs/1409.0533}{{\tt arXiv:1409.0533 [gr-qc]}}.
%%CITATION = ARXIV:1409.0533;%%.

\bibitem{HorSan14}
G.~T. Horowitz and J.~E. Santos, ``{Geons and the Instability of Anti-de Sitter
  Spacetime},''
\href{http://arxiv.org/abs/1408.5906}{{\tt arXiv:1408.5906 [gr-qc]}}.
%%CITATION = ARXIV:1408.5906;%%.

\bibitem{Mal12}
M.~Maliborski, ``{Instability of Flat Space Enclosed in a Cavity},''
  \href{http://dx.doi.org/10.1103/PhysRevLett.109.221101}{{\em Phys.Rev.Lett.}
  {\bf 109} (2012)  221101},
\href{http://arxiv.org/abs/1208.2934}{{\tt arXiv:1208.2934 [gr-qc]}}.
%%CITATION = ARXIV:1208.2934;%%.

\bibitem{BucLeh12}
A.~Buchel, L.~Lehner, and S.~L. Liebling, ``{Scalar Collapse in AdS},''
  \href{http://dx.doi.org/10.1103/PhysRevD.86.123011}{{\em Phys.Rev.} {\bf D86}
  (2012)  123011},
\href{http://arxiv.org/abs/1210.0890}{{\tt arXiv:1210.0890 [gr-qc]}}.
%%CITATION = ARXIV:1210.0890;%%.

\bibitem{BizJal13}
P.~Bizon and J.~Jalmuzna, ``{Globally regular instability of $AdS_3$},''
  \href{http://dx.doi.org/10.1103/PhysRevLett.111.041102}{{\em Phys.Rev.Lett.}
  {\bf 111} (2013) no.~4, 041102},
\href{http://arxiv.org/abs/1306.0317}{{\tt arXiv:1306.0317 [gr-qc]}}.
%%CITATION = ARXIV:1306.0317;%%.

\bibitem{Jal13}
J.~Jalmuzna, ``{Three-dimensional gravity and instability of $AdS_{3}$},''
\href{http://arxiv.org/abs/1311.7409}{{\tt arXiv:1311.7409 [gr-qc]}}.
%%CITATION = ARXIV:1311.7409;%%.

\bibitem{BizRos14}
P.~Bizon and A.~Rostworowski, ``{Comment on "Holographic Thermalization,
  stability of AdS, and the Fermi-Pasta-Ulam-Tsingou paradox" by V.
  Balasubramanian et al},''
\href{http://arxiv.org/abs/1410.2631}{{\tt arXiv:1410.2631 [gr-qc]}}.
%%CITATION = ARXIV:1410.2631;%%.

\bibitem{DFLY14}
F.~V. Dimitrakopoulos, B.~Freivogel, M.~Lippert, and I.-S. Yang, ``{Instability
  corners in AdS space},''
\href{http://arxiv.org/abs/1410.1880}{{\tt arXiv:1410.1880 [hep-th]}}.
%%CITATION = ARXIV:1410.1880;%%.

\bibitem{CraEvn14}
B.~Craps, O.~Evnin, and J.~Vanhoof, ``{Renormalization group, secular term
  resummation and AdS (in)stability},''
\href{http://arxiv.org/abs/1407.6273}{{\tt arXiv:1407.6273 [gr-qc]}}.
%%CITATION = ARXIV:1407.6273;%%.

\bibitem{CraEvn14a}
B.~Craps, O.~Evnin, and J.~Vanhoof, ``{Renormalization, averaging, conservation
  laws and AdS (in)stability},''
\href{http://arxiv.org/abs/1412.3249}{{\tt arXiv:1412.3249 [gr-qc]}}.
%%CITATION = ARXIV:1412.3249;%%.

\bibitem{BucGre14}
A.~Buchel, S.~R. Green, L.~Lehner, and S.~L. Liebling, ``{Conserved quantities
  and dual turbulent cascades in Anti-de Sitter spacetime},''
\href{http://arxiv.org/abs/1412.4761}{{\tt arXiv:1412.4761 [gr-qc]}}.
%%CITATION = ARXIV:1412.4761;%%.

\bibitem{BizMal15}
P.~Bizon, M.~Maliborski, and A.~Rostworowski, ``{Resonant dynamics and the
  instability of anti-de Sitter spacetime},''
\href{http://arxiv.org/abs/1506.03519}{{\tt arXiv:1506.03519 [gr-qc]}}.
%%CITATION = ARXIV:1506.03519;%%.

\bibitem{BucGre15}
A.~Buchel, S.~R. Green, L.~Lehner, and S.~L. Liebling, ``{Reply to "Comment on
  two-mode stability islands around AdS"},''
\href{http://arxiv.org/abs/1506.07907}{{\tt arXiv:1506.07907 [gr-qc]}}.
%%CITATION = ARXIV:1506.07907;%%.

\bibitem{BucGre14a}
A.~Buchel, S.~R. Green, L.~Lehner, and S.~L. Liebling, ``{Universality of
  non-equilibrium dynamics of CFTs from holography},''
\href{http://arxiv.org/abs/1410.5381}{{\tt arXiv:1410.5381 [hep-th]}}.
%%CITATION = ARXIV:1410.5381;%%.

\bibitem{BasKri15}
P.~Basu, C.~Krishnan, and P.~Bala~Subramanian, ``{AdS (In)stability: Lessons
  From The Scalar Field},''
  \href{http://dx.doi.org/10.1016/j.physletb.2015.05.009}{{\em Phys.Lett.} {\bf
  B746} (2015)  261--265},
\href{http://arxiv.org/abs/1501.07499}{{\tt arXiv:1501.07499 [hep-th]}}.
%%CITATION = ARXIV:1501.07499;%%.

\bibitem{Yan15}
I.-S. Yang, ``{Missing top of the AdS resonance structure},''
  \href{http://dx.doi.org/10.1103/PhysRevD.91.065011}{{\em Phys.Rev.} {\bf D91}
  (2015) no.~6, 065011},
\href{http://arxiv.org/abs/1501.00998}{{\tt arXiv:1501.00998 [hep-th]}}.
%%CITATION = ARXIV:1501.00998;%%.

\bibitem{DimYan15}
F.~Dimitrakopoulos and I.-S. Yang, ``{Occasionally Extended Validity of
  Perturbation Theory: Persistence of AdS Stability Islands},''
\href{http://arxiv.org/abs/1507.02684}{{\tt arXiv:1507.02684 [hep-th]}}.
%%CITATION = ARXIV:1507.02684;%%.

\bibitem{GreMai15}
S.~R. Green, A.~Maillard, L.~Lehner, and S.~L. Liebling, ``{Islands of
  stability and recurrence times in AdS},''
  \href{http://dx.doi.org/10.1103/PhysRevD.92.084001}{{\em Phys. Rev.} {\bf
  D92} (2015) no.~8, 084001},
\href{http://arxiv.org/abs/1507.08261}{{\tt arXiv:1507.08261 [gr-qc]}}.
%%CITATION = ARXIV:1507.08261;%%.

\bibitem{EvnKri15}
O.~Evnin and C.~Krishnan, ``{A Hidden Symmetry of AdS Resonances},''
  \href{http://dx.doi.org/10.1103/PhysRevD.91.126010}{{\em Phys. Rev.} {\bf
  D91} (2015) no.~12, 126010},
\href{http://arxiv.org/abs/1502.03749}{{\tt arXiv:1502.03749 [hep-th]}}.
%%CITATION = ARXIV:1502.03749;%%.

\bibitem{CraEvn15a}
B.~Craps, O.~Evnin, and J.~Vanhoof, ``{Ultraviolet asymptotics and singular
  dynamics of AdS perturbations},''
  \href{http://dx.doi.org/10.1007/JHEP10(2015)079}{{\em JHEP} {\bf 10} (2015)
  079},
\href{http://arxiv.org/abs/1508.04943}{{\tt arXiv:1508.04943 [gr-qc]}}.
%%CITATION = ARXIV:1508.04943;%%.

\bibitem{CraEvn15b}
B.~Craps, O.~Evnin, P.~Jai-akson, and J.~Vanhoof, ``{Ultraviolet asymptotics
  for quasiperiodic AdS$_{4}$ perturbations},''
  \href{http://dx.doi.org/10.1007/JHEP10(2015)080}{{\em JHEP} {\bf 10} (2015)
  080},
\href{http://arxiv.org/abs/1508.05474}{{\tt arXiv:1508.05474 [gr-qc]}}.
%%CITATION = ARXIV:1508.05474;%%.

\bibitem{BanDou98}
T.~Banks, M.~R. Douglas, G.~T. Horowitz, and E.~J. Martinec, ``{AdS dynamics
  from conformal field theory},''
\href{http://arxiv.org/abs/hep-th/9808016}{{\tt arXiv:hep-th/9808016
  [hep-th]}}.
%%CITATION = HEP-TH/9808016;%%.

\bibitem{KAM}
H.~W. Broer, ``{KAM theory: The legacy of Kolmogorov's 1954 paper},'' {\em
  Bulletin of the American Mathematical Society} {\bf 41} (2004)  .

\bibitem{Biz13}
P.~Bizon, ``{Is AdS stable?},''
  \href{http://dx.doi.org/10.1007/s10714-014-1724-0}{{\em Gen.Rel.Grav.} {\bf
  46} (2014)  1724},
\href{http://arxiv.org/abs/1312.5544}{{\tt arXiv:1312.5544 [gr-qc]}}.
%%CITATION = ARXIV:1312.5544;%%.

\bibitem{Kolmogorov}
A.~C. Newell, S.~Nazarenko, and L.~Biven, ``Wave turbulence and
  intermittency,'' {\em Physica D: Nonlinear Phenomena} {\bf 152–153} (2001)
  520 -- 550.

\bibitem{WeakTurbulence}
V.~E. Zakharov, V.~S. L'vov, and G.~Falkovich, ``{Statistical Description of
  Weak Wave Turbulence},''
  \href{http://dx.doi.org/10.1007/978-3-642-50052-7_3}{{\em Springer Series in
  Nonlinear Dynamics} (1992)  63--82}.

\bibitem{HamKab06}
A.~Hamilton, D.~N. Kabat, G.~Lifschytz, and D.~A. Lowe, ``{Holographic
  representation of local bulk operators},''
  \href{http://dx.doi.org/10.1103/PhysRevD.74.066009}{{\em Phys.Rev.} {\bf D74}
  (2006)  066009},
\href{http://arxiv.org/abs/hep-th/0606141}{{\tt arXiv:hep-th/0606141
  [hep-th]}}.
%%CITATION = HEP-TH/0606141;%%.

\bibitem{HamKab06a}
A.~Hamilton, D.~N. Kabat, G.~Lifschytz, and D.~A. Lowe, ``{Local bulk operators
  in AdS/CFT: A Holographic description of the black hole interior},''
  \href{http://dx.doi.org/10.1103/PhysRevD.75.106001,
  10.1103/PhysRevD.75.129902}{{\em Phys.Rev.} {\bf D75} (2007)  106001},
\href{http://arxiv.org/abs/hep-th/0612053}{{\tt arXiv:hep-th/0612053
  [hep-th]}}.
%%CITATION = HEP-TH/0612053;%%.

\bibitem{EvnNiv15}
O.~Evnin and R.~Nivesvivat, ``{AdS perturbations, isometries, selection rules
  and the Higgs oscillator},''
\href{http://arxiv.org/abs/1512.00349}{{\tt arXiv:1512.00349 [hep-th]}}.
%%CITATION = ARXIV:1512.00349;%%.

\bibitem{DFPY15}
F.~Dimitrakopoulos, B.~Freivogel, J.~Pedraza, and I.-S. Yang, ``{to appear},''.

\end{thebibliography}\endgroup

\end{document}